\documentclass{article}
\usepackage{spconf}
\usepackage{graphicx}
\usepackage{bm}
\usepackage{cite}
\usepackage{amsmath,amssymb,amsfonts,mathrsfs}
\usepackage{bbm}
\usepackage{amsxtra}
\usepackage{threeparttable}
\usepackage{algorithm,algorithmic}
\usepackage{etoolbox,siunitx}
\usepackage{multirow, booktabs}
\usepackage{balance}
\usepackage{textcomp}
\usepackage{xcolor}
\usepackage{url}
\usepackage{pifont}
\usepackage{threeparttable}
\usepackage{hyperref}
\usepackage{makecell}
\usepackage{setspace}
\usepackage{colortbl}
\robustify\bfseries

\newcommand{\cmark}{\textcolor{green}{\ding{51}}}
\newcommand{\xmark}{\textcolor{red}{\ding{55}}}

\title{FlexIO: Flexible Single- and Multi-Channel \\ Speech Separation and Enhancement}
\name{{\shortstack[c]{
      Yoshiki Masuyama$^{1}$,
      Kohei Saijo$^{2}$,
      Francesco Paissan$^{1,3}$,
      Jiangyu Han$^{4}$,
      Marc Delcroix$^{5}$, \\
      Ryo Aihara$^{1}$,
      Fran\c{c}ois G.\ Germain$^{1}$,
      Gordon Wichern$^{1}$,
      Jonathan Le Roux$^{1}$%
      \thanks{This work was done while F. Paissan was an intern at MERL.}
}}}
\address{$^{1}$Mitsubishi Electric Research Laboratories (MERL), Cambridge, USA \\
$^{2}$Waseda University, Tokyo, Japan \;
$^{3}$University of Trento, Trento, Italy \\ 
$^{4}$Brno University of Technology, Brno, Czechia \;
$^{5}$NTT, Inc., Kyoto, Japan \;
}
\begin{document}
\ninept
\setlength{\abovedisplayskip}{6pt}
\setlength{\belowdisplayskip}{6pt}
\maketitle
\begin{abstract}
Speech separation and enhancement (SSE) has advanced remarkably and achieved promising results in controlled settings, such as a fixed number of speakers and a fixed array configuration.
Towards a universal SSE system, single-channel systems have been extended to deal with a variable number of speakers (i.e., outputs).
Meanwhile, multi-channel systems accommodating various array configurations (i.e., inputs) have been developed.
However, these attempts have been pursued separately.
In this paper, we propose a flexible input and output SSE system, named FlexIO.
It performs conditional separation using prompt vectors, one per speaker as a condition, allowing separation of an arbitrary number of speakers.
Multi-channel mixtures are processed together with the prompt vectors via an array-agnostic channel communication mechanism.
Our experiments demonstrate that FlexIO successfully covers diverse conditions with one to five microphones and one to three speakers.
We also confirm the robustness of FlexIO on CHiME-4 real data.
\end{abstract}

\section{Introduction}
\label{sec:intro}

Speech separation and enhancement (SSE) aims to isolate individual speech signals from mixtures contaminated by background noise and reverberation~\cite{Gannot2017,Wang2018review}.
SSE is crucial not only for improving speech quality but also as a front-end for speech recognition~\cite{Neumann2020,Raj2021,Masuyama2026}.
With the advent of deep learning, SSE has shown remarkable results even under single-channel conditions~\cite{Hershey2016,Kolbaek2017}.
Dual-path modeling~\cite{Luo2020dprnn} in the short-time Fourier transform (STFT) domain, in particular, has yielded promising performance~\cite{Yang2022,Saijo2024,Wang2023tfgridnet}.
This approach has also been used in multi-channel SSE systems by concatenating the channel-wise STFT coefficients as input~\cite{Wang2023tfgridnet,Kalkhorani2024}.
These studies, however, mainly focused on matched conditions, i.e., the numbers of speakers and microphones are the same during training and evaluation.
A single SSE model capable of flexibly handling various numbers of microphones (input) and speakers (output) remains underexplored.

Several attempts have been made in single-channel SSE to increase the flexibility against the number of speakers in the mixture.
Recursive separation schemes extract each speaker one by one from the mixture~\cite{Kinoshita2018,Takahashi19}, and attractor-based methods internally estimate the number of speakers before separation~\cite{Chetupalli2023,Saijo2023,Dang25}.
More recently, task-aware unified source separation (TUSS) introduced prompt-conditional separation that can deal with an arbitrary number of speakers depending on the number of given prompt vectors~\cite{Saijo2025}.
Prompt-conditional separation allows users to explicitly specify the number of output streams by changing the number of prompt vectors, and this controllability is beneficial if the number of speakers is known or estimated in advance.

In multi-channel SSE, pioneering work was based on array signal processing~\cite{Hiroe2006,Souden2010} and has been combined with single-channel neural networks, e.g., in mask-based beamforming~\cite{Heymann2016,Erdogan2016}.
Mask-based beamforming has been widely used due to its flexibility with regard to the number of microphones and the array geometry.
Recent attempts solely adopt neural networks since the traditional array signal processing can be a performance bottleneck~\cite{Wang2020taslp,Wang2023tfgridnet,Kalkhorani2024}.
A na\"ive implementation is to concatenate the channel-wise STFT coefficients as the network input~\cite{Wang2023tfgridnet,Kalkhorani2024}, but this cannot accommodate mixtures with different numbers of channels.
To overcome this limitation, array-agnostic channel communication mechanisms have been developed~\cite{Luo2020tac,Pandey2022}, which deal with an arbitrary number of microphones and exchange the information across channels.
Especially, the transform-average-concatenate (TAC) mechanism~\cite{Luo2020tac} has been successfully applied to universal speech enhancement~\cite{Zhang2023}.

\begin{table}[t]
    \centering
    \caption{
    Comparison of FlexIO and flexible SSE systems, where $M$ and $N$ are the number of microphones and speakers, respectively.
    Controllability indicates whether users can specify the number of output streams during inference.
    }
    \resizebox{\linewidth}{!}{
    \begin{tabular}{ccccc}
        \toprule
        & Flexible $M$ & Flexible $N$ & Controllability \\
        \midrule
        TPARN~\cite{Pandey2022}, USES~\cite{Zhang2023} & \cmark & \xmark & \xmark \\
        SepEDA~\cite{Chetupalli2023}, TUSS~\cite{Saijo2025} & \xmark & \cmark & \cmark \\
        VarArray~\cite{Yoshioka2022} & \cmark & \cmark & \xmark \\
        DNN-IVA~\cite{Scheibler2021} & \phantom{$^\ddagger$}\cmark$^\dagger$ & \cmark & \cmark \\
        \midrule
        FlexIO (Proposed) & \cmark & \cmark & \cmark \\
        \bottomrule
    \end{tabular}
    }
    \begin{tablenotes}[flushleft]\footnotesize
    \item[*] $^\dagger$DNN-IVA is not applicable to single-channel scenarios.
    \end{tablenotes}
    \label{tab:sales}
    \vskip -4mm
\end{table}

Prior works have typically tackled the flexibility with regard to either the number of microphones $M$ or the number of speakers $N$ in isolation, as summarized in Table~\ref{tab:sales}.
Towards a more flexible SSE system, VarArray~\cite{Yoshioka2022} exploits TAC and supports a variable number of speakers in a continuous speech separation scheme. However, the number of concurrent output streams is fixed, and the model does not exploit prior knowledge about the expected number of speakers.
Independent vector analysis with neural source modeling (DNN-IVA)~\cite{Scheibler2021} offers both flexibility and controllability, but its performance is limited in underdetermined scenarios, i.e., $N > M$, due to its dependence on array signal processing.

\begin{figure}[t!]
\centering
\includegraphics[width=0.85\columnwidth]{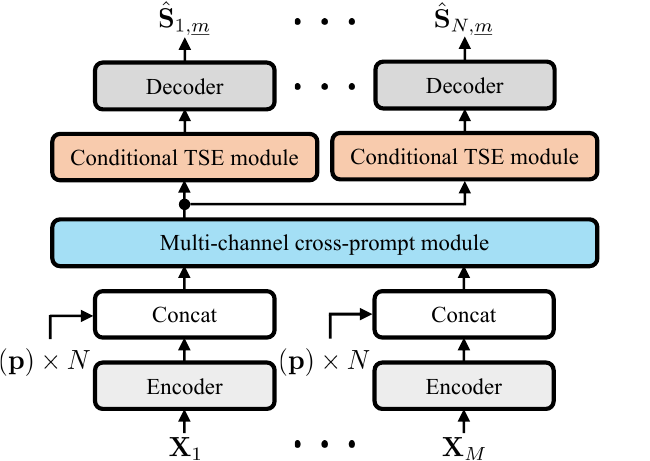}
\vskip -2mm
\caption{
FlexIO takes $N$ prompt vectors $\mathbf{p}$ and an $M$-channel mixture $(\mathbf{X}_1, \ldots, \mathbf{X}_M)$ as input, and isolates each speaker at a reference channel $\underline{m}$.
Modules of the same color share parameters.
}
\label{fig:overview}
\vskip -3mm
\end{figure}

In this paper, we propose FlexIO, a flexible and versatile SSE system that can handle an arbitrary combination of $M$ and $N$.
FlexIO extends the single-channel TUSS framework~\cite{Saijo2025} to support multi-channel input in an array-agnostic manner as depicted in Fig.~\ref{fig:overview}.
The multi-channel cross-prompt module processes the mixture representation and the given prompt vectors using time-frequency dual-path modeling with channel communication mechanisms.
Then, we perform target speaker extraction (TSE) based on the processed mixture representation and prompt vectors at the reference channel.
Our experiments on comprehensive scenarios demonstrate that FlexIO successfully copes with various input and output conditions with strong performance.

\vspace{-.25cm}
\section{Preliminaries}
\vspace{-.15cm}
\subsection{Problem settings}

Let us consider the STFT of an $M$-channel reverberant noisy mixture of $N$ speakers, $(\mathbf{X}_1, \ldots, \mathbf{X}_M)$, where $\mathbf{X}_m \in \mathbb{C}^{T \times F}$, and $T$ and $F$ denote the number of time frames and frequency bins, respectively.
Our goal is to estimate the anechoic source image of each speaker regardless of the number of microphones and speakers:
\begin{equation}
\{\hat{\mathbf{S}}_{1, \underline{m}}, \dots, \hat{\mathbf{S}}_{N, \underline{m}}\} \leftarrow \mathrm{SSE}(\mathbf{X}_1, \ldots, \mathbf{X}_M),
\label{eq:sse}
\end{equation}
where $\mathbf{S}_{n, \underline{m}} \in \mathbb{C}^{T \times F}$ denotes the desired source image of the $n$-th speaker at a reference channel $\underline{m}$.
The number of speakers $N$ can be different for each mixture, but we assume it is estimated in advance.

\vspace{-.15cm}
\subsection{Single-channel TUSS for SSE}

We review the application to SSE of TUSS~\cite{Saijo2025}, a prompt-conditional separation system on top of which we build our proposed FlexIO.
In this section, we omit the microphone index $m$ as TUSS was developed as a single-channel system.
First, we encode the real and imaginary parts of the mixture $\mathbf{X}$ into an initial mixture representation $\mathbf{Z} \in \mathbb{R}^{D \times T \times F}$ by applying 2D convolution and global layer normalization, where $D$ is the feature dimension.
Then, to extract $N$ speakers, we concatenate $N$ repeated prompt vectors $\mathbf{p} \!\in\! \mathbb{R}^D$ to the mixture representation for subsequent prompt-aware processing, leading to the representation $\mathbf{Z}' \!=\! [\mathbf{P}_1, \ldots, \mathbf{P}_N, \mathbf{Z}] \in \mathbb{R}^{D \times T' \times F}$, where $\mathbf{P}_n$ is an $F$ repeat of $\mathbf{p}$ regardless of $n$, and $T' = T+N$.
We share the same initial prompt vectors across speakers to handle an arbitrary number of speakers, but they will be updated to capture the characteristics of different speakers in the following module.

A key component of TUSS is the cross-prompt module that updates the mixture representation and prompt vectors simultaneously to capture the relation between the mixture and each speaker.
This module applies multiple TF-locoformer blocks~\cite{Saijo2024}, time-frequency dual-path modeling blocks with multi-head self-attention (MHSA), to the concatenated representation $\mathbf{Z}'$.
In the temporal modeling, $\mathbf{Z}'$ is viewed as $F$ separate sequences of $D$-dimensional features with length $T'$.
Then, MHSA and 1D convolution with a SwiGLU activation~\cite{Saijo2024} are applied along the time axis:
\begin{align}
\mathbf{Z}' &\leftarrow \mathbf{Z}' + \mathrm{ConvSwiGLU}(\mathbf{Z}'), \label{eq:locoformer-conv1} \\
\mathbf{Z}' &\leftarrow \mathbf{Z}' + \mathrm{MHSA}(\mathrm{Norm}(\mathbf{Z}')), \label{eq:locoformer-mhsa} \\
\mathbf{Z}' &\leftarrow \mathbf{Z}' + \mathrm{ConvSwiGLU}(\mathbf{Z}'), \label{eq:locoformer-conv2}
\end{align}
where $\mathrm{Norm}$ is the root mean square group normalization~\cite{Saijo2024}.
Thanks to the rotary positional encoding~\cite{Su2024} in MHSA, the identical prompt vectors are updated to different values.
The frequency modeling performs similar operations along the frequency axis at each time frame.

The subsequent conditional TSE module isolates speaker-wise representations.
The output $\tilde{\mathbf{Z}}'$ of the cross-prompt module is split into components corresponding to each prompt and the mixture: $(\tilde{\mathbf{P}}_1, \ldots, \tilde{\mathbf{P}}_N, \tilde{\mathbf{Z}}) \leftarrow \tilde{\mathbf{Z}}'$, where $\tilde{\mathbf{P}}_n$ is expected to capture the characteristics of each speaker.
The speaker-wise representation is then obtained via a Hadamard product following the conventional TSE framework~\cite{Vzmolikova2019}: $\tilde{\mathbf{Z}}_n = \tilde{\mathbf{P}}_n \odot\ \tilde{\mathbf{Z}} \in \mathbb{R}^{D \times T \times F}$, where $\tilde{\mathbf{P}}_n$ is broadcasted along the time axis.
We then separately refine $\tilde{\mathbf{Z}}_n$ for each speaker using additional TF-locoformer blocks, where the model parameters are shared across speakers.
The output of the conditional TSE module is fed to a 2D deconvolution decoder and multiplied by the STFT of the mixture $\mathbf{X}$ as complex time-frequency masks%
\footnote{
We use 2D deconvolution to predict time-frequency masks instead of band-split processing used in~\cite{Saijo2025}, since we focus only on 16 kHz speech.
}~\cite{Williamson2016}.
TUSS can deal with a variable number of speakers, including a single speaker, by adjusting the number of prompt vectors, although its input is limited to a single channel.

\vspace{-.25cm}
\section{FlexIO}
\vspace{-.15cm}
\subsection{Overview of FlexIO}

To realize flexible multi-channel SSE, FlexIO integrates a channel communication mechanism into the cross-prompt module of TUSS, and its entire architecture is designed to generalize to diverse array configurations as depicted in Fig.~\ref{fig:overview}.
First, each channel component of the $M$-channel mixture $(\mathbf{X}_1, \ldots, \mathbf{X}_M)$ is separately encoded.
The multi-channel cross-prompt module handles the $M$-channel features $(\mathbf{Z}'_1, \ldots, \mathbf{Z}'_M)$ separately, except for the additional channel communication mechanism: TAC~\cite{Luo2020tac},  cross-channel attention~\cite{Pandey2022}, or co-attention~\cite{Horiguchi2022}.
Then, we take the processed representations at the
reference channel and feed them into the conditional TSE module: $\tilde{\mathbf{Z}}_{n,\underline{m}}  = \tilde{\mathbf{P}}_{n,\underline{m}} \odot\ \tilde{\mathbf{Z}}_{\underline{m}}$.
We expect this to encourage FlexIO to produce estimates aligned with the ground truth at the reference channel~\cite{Yoshioka2022,Zhang2023}.
Unlike existing systems based on array signal processing~\cite{Scheibler2021}, FlexIO can handle even single-channel mixtures due to its careful design.
In addition, inheriting the prompt-conditional separation of TUSS, FlexIO can deal with a variable number of speakers by adjusting the number of prompt vectors.
With one prompt vector, FlexIO focuses on suppressing background noise and reverberation, thus covering both enhancement and separation in a unified manner.

\subsection{TAC and cross-channel attention}

We require a channel communication mechanism to exchange information across channels while keeping the entire architecture array-agnostic.
TAC is a well-developed channel communication mechanism~\cite{Luo2020tac}, and we adopt it after each TF-locoformer block.
In detail, we separately encode each TF-bin to a hidden dimension $E$:
\begin{equation}
    \mathbf{W}_{m} \leftarrow \mathrm{FC}_\mathrm{in}(\mathbf{Z}_{m}') \in \mathbb{R}^{E\times T' \times F},
\end{equation}
where FC denotes a fully-connected layer with PReLU activation.
The projected representation is averaged over channels and fed into another fully-connected layer:
\begin{equation}
    \underline{\mathbf{W}} \leftarrow \mathrm{FC}_\mathrm{avg} \bigg( \frac{1}{M} \sum_{m=1}^M \mathbf{W}_{m} \bigg) \in \mathbb{R}^{E\times T' \times F}.
\end{equation}
The global representation $\underline{\mathbf{W}}$ is concatenated with the channel-wise representation $\mathbf{W}_{m}$ and projected back to the original dimension with a skip connection:
\begin{equation}
    \mathbf{Z}_{m}' \leftarrow \mathbf{Z}_{m}' + \mathrm{Norm}(\mathrm{FC}_\mathrm{cat}([\mathbf{W}_{m}; \underline{\mathbf{W}}])).
\end{equation}
This mechanism can handle an arbitrary number of microphones due to the nature of average pooling.
In addition, it falls back to standard fully-connected layers in single-channel scenarios.

Another channel communication mechanism is cross-channel attention~\cite{Pandey2022,Wang2022cca}.
It applies MHSA across channels without positional encoding, which enables generalization to a different number of microphones.
The cross-channel attention realizes a dynamic weighted average, while TAC aggregates the representation with the same weight for each channel regardless of the given mixture.

\vspace{-0.2cm}
\subsection{Channel communication via co-attention}

We also explore the co-attention mechanism, originally proposed for diarization~\cite{Horiguchi2022}.
In contrast to TAC and cross-channel attention, co-attention does not introduce additional layers but tweaks existing channel-wise MHSA.
Specifically, the original channel-wise MHSA for temporal modeling in \eqref{eq:locoformer-mhsa} can be realized by 
\begin{align}
    \mathbf{A}_{m,f}^{(h)} &\leftarrow \mathrm{softmax}\left( \frac{\mathbf{Q}_{m,f}^{(h)} \mathbf{K}_{m,f}^{(h)\mathsf{T}}}{\sqrt{D_\mathrm{att}}} \right) \in \mathbb{R}^{T' \times T'}, \label{eq:ch-wise-att} \\
    \!\!\mathbf{U}_{m,f} &\leftarrow \mathrm{FC}_\mathrm{att}\left([\mathbf{A}_{m,f}^{(1)}\mathbf{V}_{m,f}^{(1)}; \cdots; \mathbf{A}_{m,f}^{(H)}\mathbf{V}_{m,f}^{(H)}] \right)\!\in \mathbb{R}^{D \times T'}\!,\!\!
    \label{eq:head-agg}
\end{align}
where $H$ is the number of attention heads, and  $\mathbf{Q}_{m,f}^{(h)}$, $\mathbf{K}_{m,f}^{(h)}$, and $\mathbf{V}_{m,f}^{(h)}$ respectively are the query, key, and value for the $h$-th head at the $f$-th frequency.
They are calculated from $\mathbf{Z}'$ after the normalization using fully-connected layers.
In \eqref{eq:head-agg}, $\mathrm{FC}_\mathrm{att}$ aggregates the outputs of multiple heads, and the output $\mathbf{U}_{m,f}$ is stacked over all frequencies.
The co-attention mechanism replaces the channel-wise weight in \eqref{eq:ch-wise-att} with the following channel-invariant weight~\cite{Horiguchi2022}:
\begin{equation}
    \mathbf{A}_{f}^{(h)} \leftarrow \mathrm{softmax}\left( \frac{\sum_{m=1}^M \mathbf{Q}_{m,f}^{(h)} \mathbf{K}_{m,f}^{(h)\mathsf{T}}}{\sqrt{D_\mathrm{att}M}} \right) \in \mathbb{R}^{T' \times T'},
    \label{eq:co-attention}
\end{equation}
and $\mathbf{A}_{f}^{(h)}$ is used at all channels.
The additional factor $\sqrt{M}$ arises from interpreting the numerator of \eqref{eq:co-attention} as the concatenation of $M$ keys and queries.
While co-attention does not explicitly exchange the representation across channels, it can improve multi-channel SSE performance as shown in the following experiments.

\vspace{-.2cm}
\section{Experiments}
\label{sec:experiment}
\vspace{-.1cm}
\subsection{Datasets}
\vspace{-.05cm}

\begin{table}[t]
    \centering
    \vspace{-6pt}
    \caption{
    Datasets used in our SSE experiments. ``A'' and ``R'' in the condition denote anechoic and reverberant settings.
    The 3- and 5-channel data are excluded from the training and validation sets. 
    }
    \resizebox{\linewidth}{!}
    {\setlength{\tabcolsep}{4pt}
    \begin{tabular}{rccc}
        \toprule
        & \#Ch $M$ & \#Spks $N$ & Acoustic condition \\
        \midrule
        CHiME-4~\cite{Vincent2017} & $\{1, 2, 4\}$, $\{3, 5\}$  & 1 & Noisy A \\
        WSJ0-mix~\cite{Hershey2016}\phantom{3} & 1 & $\{2, 3\}$ & Clean \\
        WHAM!~\cite{Wichern2019} & 1 & $\{1, 2\}$ & Noisy A \\
        WHAMR!~\cite{Maciejewski2020} & $\{1, 2\}$ & $\{1, 2\}$ & Noisy A/R \\
        WSJ1-CHiME~\cite{Scheibler2021} & $\{2, 4\}$, 3 & $\{2, 3\}$ & Noisy R \\
        \bottomrule
    \end{tabular}
    }
    \label{tab:datasets}
    \vskip -4mm
\end{table}

To validate the flexibility of our proposed FlexIO, we combined several SSE datasets, as summarized in Table~\ref{tab:datasets}.
The CHiME-4 dataset provides not only simulated but also real 6-channel noisy recordings of a single speaker~\cite{Vincent2017}.
WSJ0-mix is a standard speech separation dataset~\cite{Hershey2016}, and we further incorporated its noisy (WHAM!~\cite{Wichern2019}) and noisy reverberant (WHAMR!~\cite{Maciejewski2020}) extensions to cover diverse acoustic conditions.
In addition, our training set includes WSJ1-CHiME developed in \cite{Scheibler2021}, which covers various numbers of microphones and speakers%
\footnote{\url{https://github.com/fakufaku/create_wsj1_2345_db}}.
Finally, our training dataset covers 1 to 3 speakers with 1 to 4 channels, where the sampling rate was set to 16 kHz for all recordings.
The 3- and 5-channel recordings were excluded from the training and validation sets to assess the generalization capability under unseen conditions.

\begin{table*}[t]
\centering
\sisetup{
detect-weight,
mode=text,
tight-spacing=true,
round-mode=places,
round-precision=1,
table-format=2.1,
table-number-alignment=center
}
    \vspace{-6pt}
\caption{Speech enhancement performance under diverse conditions.
The numbers in parentheses denote the numbers of speakers and microphones, i.e., ($N$-$M$).
For the WHAMR! dataset, ``A'' and ``R'' indicate the anechoic and reverberant conditions, respectively.
``Comm.'' shows the channel communication mechanism, where ``1ch'' indicates that speech enhancement is performed solely on the reference channel.
The best and second best results are highlighted in blue and orange, respectively
}
\label{table:se-results}
\resizebox{\linewidth}{!}{
\setlength{\tabcolsep}{4pt}
\begin{tabular}{lc*{16}{c}}
    \toprule
    & & & \multicolumn{3}{c}{WHAM! (1-1)} & \multicolumn{3}{c}{WHAMR! A (1-2)} & \multicolumn{3}{c}{WHAMR! R (1-2)} & \multicolumn{3}{c}{CHiME-4 (1-4)} & \multicolumn{3}{c}{CHiME-4 (1-5)} \\
    \cmidrule(lr){4-6} \cmidrule(lr){7-9} \cmidrule(lr){10-12} \cmidrule(lr){13-15} \cmidrule(lr){16-18} 
    & Comm. & \#Params ($10^6$) & SDR  & STOI  & PESQ & SDR  & STOI  & PESQ & SDR  & STOI  & PESQ & SDR & STOI & PESQ & SDR  & STOI  & PESQ \\
    \midrule
    USES~\cite{Zhang2023}& TAC & 3.05 & 10.2 & 85.7 & 1.65 & \cellcolor{cyan!20}15.8 & \cellcolor{cyan!20}96.4 & 2.55 & 13.8 & \cellcolor{cyan!20}96.0 & 2.51 & 18.3 & 96.6 & 2.46 & 18.3 & 97.8 & 2.95 \\
    TUSS~\cite{Saijo2025} & 1ch & 3.42 & \cellcolor{orange!20}13.6 & \cellcolor{orange!20}94.2 & \cellcolor{orange!20}2.29 & 13.6 & 94.1 & 2.28 & 12.4 & 93.7 & 2.24 & 17.1 & 95.7 & 2.31 & 17.1 & 95.7 & 2.31 \\
    \midrule
    \multirow{4}{*}{FlexIO} & TAC (M) & 3.59 & 13.5 & 94.0 & 2.25 & 15.4 & 95.6 & 2.55 & 14.2 & 95.3 & 2.50 & 19.3 & 97.3 & 2.54 & 19.6 & 97.5 & 2.60 \\
    & ChAtt (M) & 3.49 & \cellcolor{orange!20}13.6 & 94.1 & 2.25 & 15.6 & 95.9 & \cellcolor{orange!20}2.57 & \cellcolor{orange!20}14.8 & 95.7 & \cellcolor{orange!20}2.56 & 19.5 & 97.4 & 2.55 & 20.2 & 97.6 & 2.61 \\
    & CoAtt (M) & 3.42 & 13.5 & 94.0 & 2.26 & 15.5 & 95.7 & 2.56 & 14.4 & 95.4 & 2.52 & \cellcolor{orange!20}20.8 & \cellcolor{orange!20}97.8 & \cellcolor{orange!20}2.72 & \cellcolor{orange!20}21.7 & \cellcolor{orange!20}98.1 & \cellcolor{orange!20}2.81 \\
    & CoAtt (L) & 7.35 & \cellcolor{cyan!20}13.8 & \cellcolor{cyan!20}94.4 & \cellcolor{cyan!20}2.31 & \cellcolor{cyan!20}15.8 & \cellcolor{orange!20}96.0 & \cellcolor{cyan!20}2.61 & \cellcolor{cyan!20}14.9 & \cellcolor{orange!20}95.8 & \cellcolor{cyan!20}2.58 & \cellcolor{cyan!20}21.3 & \cellcolor{cyan!20}98.0 & \cellcolor{cyan!20}2.86 & \cellcolor{cyan!20}22.3 & \cellcolor{cyan!20}98.3 & \cellcolor{cyan!20}2.96 \\
    \bottomrule
\end{tabular}
}
\vspace{-5mm}
\end{table*}

\begin{table*}[t]
\centering
\sisetup{
detect-weight, %
mode=text, %
tight-spacing=true,
round-mode=places,
round-precision=1,
table-format=2.1,
table-number-alignment=center
}
\caption{Speech separation performance in various settings.
FlexIO and TUSS are the same models as those used for speech enhancement in Table~\ref{table:se-results}.
While DNN-IVA uses the same model for all conditions, TF-GridNet$^\dagger$ is separately trained for the 1- and 2-channel conditions.
}
\label{table:ss-results}
\resizebox{\linewidth}{!}{
\setlength{\tabcolsep}{4pt}
\begin{tabular}{lc*{16}{c}}
    \toprule
    & & & \multicolumn{3}{c}{WHAMR! R (2-1)} &  \multicolumn{3}{c}{WHAMR! R (2-2)} & \multicolumn{3}{c}{WSJ1-CHiME (2-2)} & \multicolumn{3}{c}{WSJ1-CHiME (2-4)} & \multicolumn{3}{c}{WSJ1-CHiME (3-3)} \\
    \cmidrule(lr){4-6} \cmidrule(lr){7-9} \cmidrule(lr){10-12} \cmidrule(lr){13-15} \cmidrule(lr){16-18}
    & Comm. & \#Params ($10^6$) & SDR  & SIR  & PESQ & SDR  & SIR  & PESQ & SDR  & SIR  & PESQ & SDR  & SIR  & PESQ & SDR  & SIR  & PESQ \\
    \midrule
    DNN-IVA~\cite{Scheibler2021} & - & 5.13 & N/A & N/A & N/A & - & - & - & 10.7 & 24.1 & - & - & - & - & \phantom{1}7.7 & 20.1 & - \\
    TF-GridNet$^\dagger$~\cite{Masuyama2026} & - & 8.38 & 9.3 & \cellcolor{cyan!20}27.0 & 1.79 & 11.7 & \cellcolor{cyan!20}29.6 & 2.20 & - & - & - & - & - & - & - & - & - \\
    TUSS~\cite{Saijo2025} & 1ch & 3.42 & \cellcolor{orange!20}9.5 & 25.3 & \cellcolor{orange!20}1.82 & \phantom{1}9.5 & 25.3 & 1.82 & 14.9 & 29.1 & 2.73 & 15.1 & 29.4 & 2.75 & 11.4 & 23.1 & 2.15 \\
    \midrule
    \multirow{4}{*}{FlexIO} & TAC (M) & 3.59 & 8.9 & 24.0 & 1.77 & 11.8 & 28.5 & 2.14 & 18.5 & 33.9 & 3.22 & 19.5 & 25.0 & 3.37 & 15.7 & 28.4 & 2.81 \\
    & ChAtt (M) & 3.49 & 9.0 & 23.7 & 1.75 & \cellcolor{orange!20}12.4 & 29.5 & \cellcolor{cyan!20}2.22 & \cellcolor{orange!20}19.5 & \cellcolor{orange!20}34.9 & \cellcolor{orange!20}3.34 & \cellcolor{orange!20}20.6 & 36.2 & \cellcolor{orange!20}3.49 & \cellcolor{orange!20}16.9 & \cellcolor{orange!20}30.1 & \cellcolor{orange!20}2.97 \\
    & CoAtt (M) & 3.42 & 9.1 & 24.4 & 1.78 & 12.1 & 29.0 & 2.18 & 18.9 & 34.7 & 3.27 & \cellcolor{orange!20}20.6 & \cellcolor{orange!20}36.4 & \cellcolor{orange!20}3.49 & 16.7 & 30.0 & 2.96 \\
    & CoAtt (L) & 7.35 & \cellcolor{cyan!20}9.7 & \cellcolor{orange!20}25.5 & \cellcolor{cyan!20}1.84 & \cellcolor{cyan!20}12.5 & \cellcolor{cyan!20}29.6 & \cellcolor{cyan!20}2.22 & \cellcolor{cyan!20}19.6 & \cellcolor{cyan!20}35.0 & \cellcolor{cyan!20}3.36 & \cellcolor{cyan!20}21.6 & \cellcolor{cyan!20}37.3 & \cellcolor{cyan!20}3.60 & \cellcolor{cyan!20}17.3 & \cellcolor{cyan!20}30.4 & \cellcolor{cyan!20}3.03 \\
    \bottomrule
\end{tabular}
}
\vspace{-4.5mm}
\end{table*}

\vspace{-.1cm}
\subsection{Model and training details}
\vspace{-.05cm}

The multi-channel cross-prompt module and the conditional TSE module consisted of $2$ and $4$ TF-locoformer blocks~\cite{Saijo2024}.
In our medium models, the configuration of each TF-locoformer block mainly followed TUSS
with medium size settings~\cite{Saijo2025}, i.e., $D=64$, $H=4$, and the kernel size and stride of the 1D convolution were $4$ and $1$, respectively, except for the temporal convolution in the multi-channel cross-prompt module.
In the multi-channel cross-prompt module, we omitted ConvSwiGLU before MHSA in \eqref{eq:locoformer-conv1} to improve the efficiency of multi-channel processing following \cite{Paissan2025}.
For large models, we set $D$ to $96$ and did not omit ConvSwiGLU in \eqref{eq:locoformer-conv1}.
Regarding the channel communication mechanism, the hidden dimension $E$ for TAC was 128, and the cross-channel attention leveraged MHSA with $4$ heads and a dimension $16$ for each head.

During training, we first sampled a pair of $N$ and $M$ for each batch.
Then, the corresponding mixtures were randomly selected and truncated at $4$ seconds.
The medium and large FlexIOs were trained up to 100 and 150 epochs, respectively, with batch sizes of 16 and 8.
Each epoch consisted of 2.5k steps.
The AdamW optimizer was used with a weight decay factor $0.01$.
The learning rate was warmed up to $0.001$ within 30k steps.
The learning rate was halved if the validation loss did not improve over 5 successive epochs, and we terminated the training when the best validation loss was not updated over 10 epochs.
The negative signal-to-noise ratio was used as the loss function with permutation invariant training~\cite{Hershey2016,Kolbaek2017}.
For evaluation, we used the signal-to-distortion ratio (SDR)~\cite{Vincent2006}, signal-to-interference ratio (SIR)~\cite{Vincent2006}, wide-band PESQ~\cite{wpesq}, and STOI~\cite{Taal2011}.

\vspace{-.1cm}
\subsection{Results}
\vspace{-.05cm}

Tables~\ref{table:se-results} and \ref{table:ss-results} respectively show enhancement and separation results under diverse conditions, where FlexIO and TUSS share the same models in both tables.
``TUSS~\cite{Saijo2025}'' corresponds to FlexIO without the channel communication mechanism, and it performs single-channel processing only on the reference channel.
In the enhancement experiment, USES~\cite{Zhang2023} is an array-agnostic system trained on multiple datasets.
To compute scores not reported in the original USES paper, we used its pre-trained model%
\footnote{\url{https://huggingface.co/espnet/Wangyou_Zhang_universal_train_enh_uses_refch0_2mem_raw}}.
When the number of microphones increased to more than one, FlexIO consistently outperforms its single-channel version (``TUSS'') regardless of the choice of the channel communication mechanism.
These results confirm that FlexIO successfully exploits spatial information in an array-agnostic manner. 
On the single-channel WHAM dataset, the models with different channel communication mechanisms result in comparable performance, slightly lagging behind the single-channel specific TUSS.
On the 2-channel WHAMR!, the cross-channel attention works marginally better than TAC and co-attention.
Meanwhile, FlexIO with co-attention performs best on the CHiME-4 dataset with more channels.
Although the model was not trained on 5-channel recordings, it successfully improves the performance when the number of microphones increases from $4$ to $5$.
This result confirms the generalization capability of FlexIO to unseen numbers of microphones.
Finally, the large FlexIO with co-attention achieves comparable or slightly better performance compared with the enhancement-specific universal model, USES~\cite{Zhang2023}.

Table~\ref{table:ss-results} summarizes the results on speech separation with various numbers of speakers.
DNN-IVA~\cite{Scheibler2021} uses the same model across different conditions, while TF-GridNet~\cite{Wang2023tfgridnet} is specific to 2-speaker cases and separately trained for 1- and 2-channel conditions following \cite{Masuyama2026}.
FlexIOs with different channel communication mechanisms show the same tendency as in the enhancement experiment.
We emphasize that FlexIO works well even on the 3-channel mixtures, even though the models were not trained on 3-channel recordings.
On WSJ1-CHiME, our FlexIO significantly outperformed DNN-IVA~\cite{Scheibler2021}.
Furthermore, the large FlexIO achieves substantially better performance than task-specific TF-GridNet under both 1- and 2-channel conditions.
These results confirm that FlexIO successfully covers diverse conditions with arbitrary numbers of microphones and speakers.

\begin{table}[t]
\sisetup{
detect-weight,
mode=text,
tight-spacing=true,
table-format=1.1,
table-number-alignment=center
}
  \vskip -2mm
  \caption{
    DNSMOS and recognition performance on CHiME-4 recordings. 
    FlexIO adopts co-attention for channel communication.
  }
  \label{tab:chime4}
  \centering
  \resizebox{\linewidth}{!}{
  \begin{tabular}{c*{2}{S}S[table-format=2.1]*{4}{S}S[table-format=2.1]}
    \toprule
     & \multicolumn{4}{c}{Simulated} & \multicolumn{4}{c}{Real} \\
     \cmidrule(lr){2-5} \cmidrule(lr){6-9} 
     & \multicolumn{2}{c}{DNSMOS} & \multicolumn{2}{c}{WER (\%)} & \multicolumn{2}{c}{DNSMOS} & \multicolumn{2}{c}{WER (\%)} \\
     \cmidrule(lr){2-3} \cmidrule(lr){4-5} \cmidrule(lr){6-7} \cmidrule(lr){8-9} 
     & {1ch} & {5ch} & {1ch} & {5ch} & {1ch} & {5ch} & {1ch} & {5ch}  \\
     \midrule 
     Noisy/MVDR & 2.08 & 2.57 & \cellcolor{cyan!20}5.8 & \cellcolor{orange!20}4.0 & 1.46 & 1.95 & \cellcolor{cyan!20}6.7 & \cellcolor{cyan!20}4.5 \\
     USES~\cite{Zhang2023} & \cellcolor{orange!20}3.03 & \cellcolor{cyan!20}3.22 & 15.2 & 4.4 & \cellcolor{cyan!20}3.07 & 1.58 & 7.4 & 85.9 \\
     U2-C~\cite{Zhang2024} & {-} & {-} & {-} & {-} & {-} & \cellcolor{cyan!20}3.08 & {-} & 10.3 \\
     FlexIO (M) & \cellcolor{cyan!20}3.11 & \cellcolor{orange!20}3.17 & \cellcolor{orange!20}6.0 & \cellcolor{cyan!20}3.9 & \cellcolor{orange!20}2.91 & \cellcolor{orange!20}3.05 & \cellcolor{orange!20}6.8 & \cellcolor{cyan!20}4.5\\
    \bottomrule
  \end{tabular}
  }
  \vskip -5mm
\end{table}

To validate the generalization capability of FlexIO, we assessed  DNSMOS (OVRL)~\cite{Reddy2022} and word error rates (WERs) on the CHiME-4 simulated and real recordings.
We used the Whisper Large v2 model for recognition following~\cite{Zhang2023,Zhang2024}.
As shown in Table~\ref{tab:chime4}, USES faces severe performance degradation on the 5-channel real data, which may be due to inconsistent SDR between channels in real recordings~\cite{Zhang2023}.
U2-C substantially mitigates this issue by carefully combining TAC and cross-channel attention~\cite{Zhang2024}.
Its WER, however, is still worse than mask-based MVDR beamforming used in \cite{Masuyama2023} and even observed signals without processing.
On the other hand, our FlexIO simultaneously achieves promising DNSMOS and WER not only on simulated but also on real recordings.

\vspace{-1mm}
\section{Conclusion}

We presented FlexIO, a versatile SSE system that flexibly deals with arbitrary numbers of microphones (inputs) and speakers (outputs).
In FlexIO, the multi-channel cross-prompt module takes a sequence of prompt vectors, controlling the number of speakers to be separated.
In addition, it exchanges information across channels in an array-agnostic manner.
Our experiments show the efficacy of FlexIO on diverse conditions.
Its extension to general audio source separation, similar to the original TUSS~\cite{Saijo2025}, and to joint source counting are directions of our future work.

\clearpage
\let\oldthebibliography\thebibliography
\renewcommand{\thebibliography}[1]{%
  \oldthebibliography{#1}%
  \footnotesize
  \setlength{\itemsep}{0.0pt}%
  \setlength{\parskip}{0.0pt}%
}
\bibliographystyle{IEEEtran}
\bibliography{strings,refs}
\end{document}